\documentclass[conference]{IEEEtran}
\IEEEoverridecommandlockouts
\usepackage[hidelinks]{hyperref}
\usepackage{orcidlink}
\usepackage{cite}
\usepackage{amsmath,amssymb,amsfonts}
\usepackage{algorithmic}
\usepackage{graphicx}
\usepackage{textcomp}
\usepackage{xcolor}
\usepackage[nameinlink]{cleveref}
\usepackage[font=small,skip=5pt]{caption}

\pdfoutput=1

\crefname{figure}{Fig.}{Fig.}
\Crefname{figure}{Fig.}{Fig.}

\graphicspath{{./img/}}

\def\BibTeX{{\rm B\kern-.05em{\sc i\kern-.025em b}\kern-.08em
    T\kern-.1667em\lower.7ex\hbox{E}\kern-.125emX}}
\begin{document}

\title{Trusting the Cloud-Native Edge: Remotely Attested Kubernetes Workers\\
\thanks{This work has been partially funded by the European Union (NATWORK project, Grant Agreement 101139285, https://natwork-project.eu/).}
}

\author{\IEEEauthorblockN{Jordi Thijsman~\orcidlink{0009-0007-2333-7051}, Merlijn Sebrechts~\orcidlink{0000-0002-4093-7338}, Filip De Turck~\orcidlink{0000-0003-4824-1199}} and Bruno Volckaert~\orcidlink{0000-0003-0575-5894}
\IEEEauthorblockA{\textit{IDLab, Department of Information Technology} \\
Ghent University - imec, Ghent, Belgium \\
jordi.thijsman@ugent.be}
}

\IEEEpubid{\makebox[\columnwidth]{Pre-print of article accepted to IEEE ICCCN 2024 \copyright2024 IEEE \hfill}
\hspace{\columnsep}\makebox[\columnwidth]{ }}

\maketitle

\begin{abstract}
A Kubernetes cluster typically consists of trusted nodes, running within the confines of a physically secure datacenter. With recent advances in edge orchestration, this is no longer the case. This poses a new challenge: how can we trust a device that an attacker has physical access to? This paper presents an architecture and open-source implementation that securely enrolls edge devices as trusted Kubernetes worker nodes. By providing boot attestation rooted in a hardware Trusted Platform Module, a strong base of trust is provided. A new custom controller directs a modified version of Keylime to cross the cloud-edge gap and securely deliver unique cluster credentials required to enroll an edge worker. The controller dynamically grants and revokes these credentials based on attestation events, preventing a possibly compromised node from accessing sensitive cluster resources. We provide both a qualitative and a quantitative evaluation of the architecture. The qualitative scenarios prove its ability to attest and enroll an edge device with role-based access control (RBAC) permissions that dynamically adjust to attestation events. The quantitative evaluation reflects an average of 10.28 seconds delay incurred on the startup time of the edge node due to attestation for a total average enrollment time of 20.91 seconds. The presented architecture thus provides a strong base of trust, securing a physically exposed edge device and paving the way for a robust and resilient edge computing ecosystem.
\end{abstract}


\section{Introduction}
As computing environments evolve, so do the challenges associated with ensuring their security. Typically, Kubernetes~\cite{Kubernetes:Applications} nodes reside within secure data centers, shielded from physical tampering and unauthorized access. However, recent advancements in edge container orchestration, such as Fledge~\cite{Goethals2022ExtendingKubelets} and KubeEdge~\cite{KubeEdge:Framework} , have ushered in a new era where clusters can extend beyond the confines of datacenters to embrace even the smallest devices. These devices are often deployed in remote and physically exposed locations making them exceptionally vulnerable to compromise.

While existing research has explored various aspects of Kubernetes security, such as protecting workloads from each other~\cite{Goel2022AuthenticatingCluster} or safeguarding them against privileged user threats~\cite{Fernandez2019SecureCloud}, our focus is on protecting the entire cluster from physically compromised nodes. By providing hardware-based boot attestation of the edge device, a trusted base can be provided that other features such as runtime attestation or kernel security features can rely on. 

Keylime~\cite{Keylime:IoT} provides an abstraction on top of TPM boot and runtime attestation, allowing for easier secure bootstrapping of managed hardware attested devices. Efforts are underway to make it easily deployable to cloud-native clusters such as Kubernetes and OpenShift. But integration of the Kubernetes API into its various parts is still lacking, making it impractical to automate attestation management. More importantly in this context, it lacks the ability to manage cluster permissions granted to an edge node in relation to attestation events. A critical aspect of our approach involves integrating flexible role-based access control (RBAC) mechanisms, to comply with Kubernetes security best practices~\cite{IslamShamim2020XIPractices}. Specifically adapted to the context of edge worker nodes, these RBAC policies enable swift and granular adjustments to access permissions in response to attestation events. These responses can prevent untrusted devices from accessing sensitive information or quarantine it to prevent scheduling of sensitive workloads. By swiftly limiting an edge node's access to cluster resources upon compromise, our architecture aims to bolster the overall security posture of distibuted Kubernetes cloud-edge clusters.

This paper presents an exploration of our proposed open-source\footnote{\href{https://github.com/idlab-discover/trust-edge}{https://github.com/idlab-discover/trust-edge}}~\cite{TrustKubernetes} platform, addressing key research questions:

\textbf{RQ1.} How can a toolchain automate enrollment of an edge device as a trusted Kubernetes worker node, ensuring full boot attestation verified by the cluster?

\textbf{RQ2.} How can such a platform integrate into the authentication and authorization framework of the orchestrator itself, in order to ensure granular permission adjustments in response to attestation events?

This paper is structured as follows: \cref{background} provides essential context on TPM attestation, measured boot, and the Keylime software and outlines other work related to these topics. \Cref{arch} delves into the components constituting the proposed architecture. \Cref{implementation} details the full enrollment and trust management process, explaining the functionalities and interactions of various components. \Cref{eval} presents the evaluation results and addresses RQ1 and RQ2. Finally, \cref{future-work} explores potential avenues for future research.

\section{Background and related work}\label{background}
\subsection{TPMs and the boot process}
A Trusted Platform Module is a secure cryptographic coprocessor, one of its most important functionalities is its ability to measure digests into so-called platform configuration registers (PCRs)~\cite{Kinney20066Registers}. The core idea of these PCRs is built around two basic operations: extend and reset. The extend operation performs an XOR operation with the current PCR state and the new digest as operands and saves the hash of the result as the new PCR value. 

Proving that a device is running trusted software is a very complex process, stretching all the way from the lowest levels of the boot sequence to the kernel, the operating system and the applications it runs. These components form a chain, relying on a lower level to tie the next to a trusted state by measuring the next binary into a TPM platform configuration register in a process called a measured boot. The lowest level of the chain is the so called core root of trust for measurement (CRTM)~\cite{CooperDavid2011SpecialGuidelines} which anchors the entire chain to a piece of immutable CPU code. If a system lacks a proper measurement for any part of this chain, it could compromise the system's ability to protect workloads as many container isolation mechanisms rely on security features provided by the Linux kernel (such as cgroups and namespaces). 

\begin{figure}[h]
    \centering
    \includegraphics[width=\linewidth]{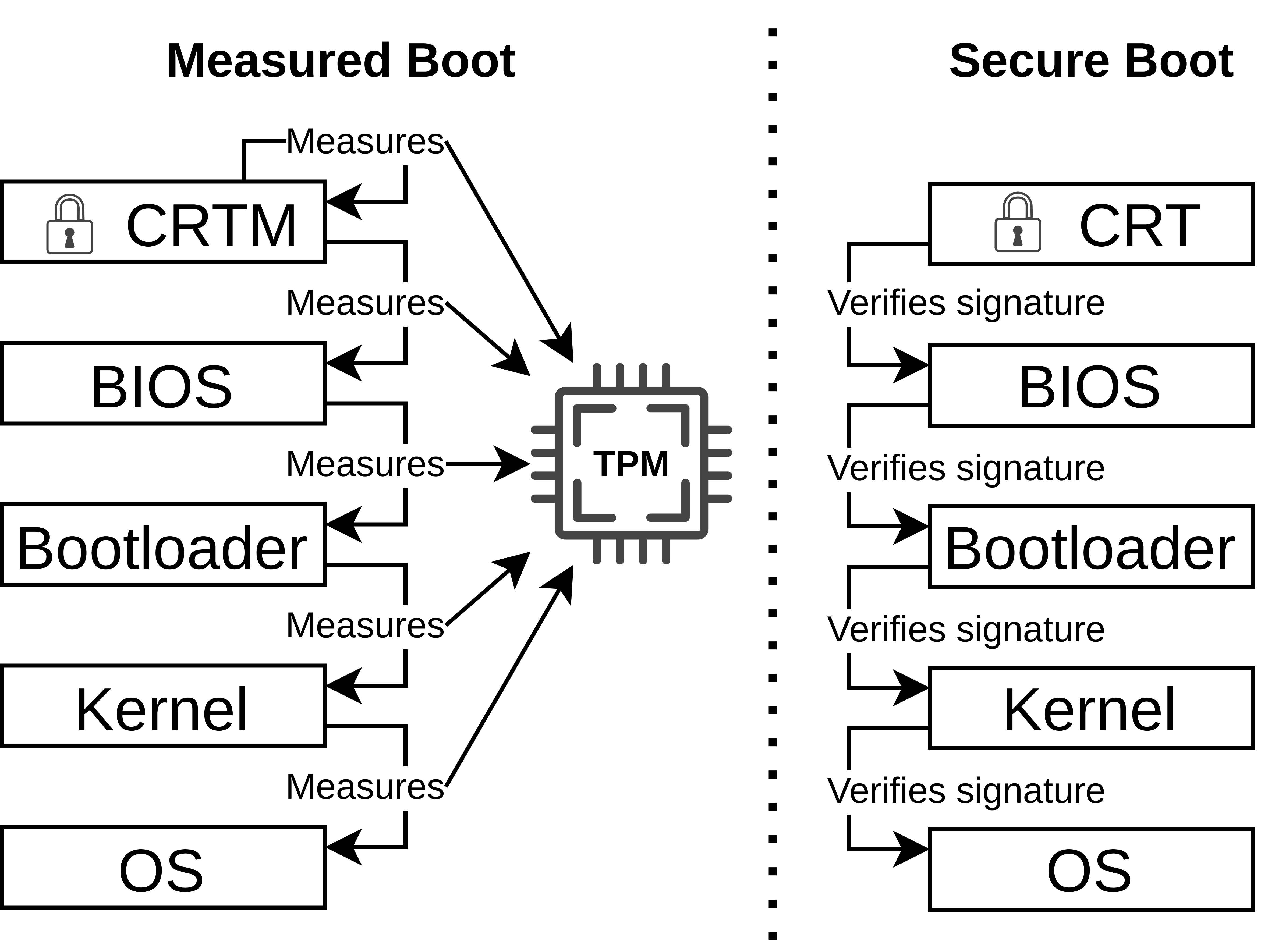}
    \caption{A measured boot extends a digest of each part of the boot chain into the TPM, this is in contrast with a secure boot which only verifies a part's signature.}
    \label{fig:tpm}
\end{figure}

It is important to recognize the distinction between such a measured boot and a secure boot. \cref{fig:tpm} provides a schematic comparison between the two. Where measured boot stores digests of each part of the boot chain in the TPM, a secure boot only validates the parts' signatures against a known good public key. If one of the system's components does not contain an approved signature, the system will simply interrupt the boot and shut down. A device using secure boot thus makes its own trust decision, compared to a measured boot, which just securely documents the boot process in PCRs and allows an external verifier to judge the TPM measurements.

TPMs are often used in edge research to provide hardware-based trust. It is, however, essential to acknowledge a critical oversight regularly present in such implementations. These designs are often implemented on devices lacking a CRTM (e.g., Raspberry Pi). Without a CRTM, the chain of trust, crucial for ensuring the integrity of the system, lacks a solid anchor  ~\cite{Arthur2015A2.0}. Consequently, reliance on a potentially compromised kernel to transmit measurements to the TPM can render the entire attestation process unreliable. Simply plugging a TPM into an edge device and trusting it is not enough. This oversight highlights the importance of ensuring that the foundational elements, such as a complete boot attestation anchored in the CRTM, are present.

This lack of manufacturer provided CRTM renders many edge devices useless in an environment where trust is required. Recent research into software TPMs could, however, provide a future for such devices. In contrast to hardware-based TPMs, software TPMs leverage Trusted Execution Environments (TEEs) to establish trust without necessitating a built-in CRTM. There are some research projects developing such TPMs~\cite{Raj2016FTPM:Chip}~\cite{Sun2018ETPM:Technology} for various platforms such as ARM TrustZone and Intel SGX, however no implementation has been widely adopted.

A final issue with TPMs is their complexity. Interacting with a TPM requires a very good understanding of the complex architecture~\cite{2019TrustedArchitecture} and is generally done by transmitting low level binary commands~\cite{2019TrustedCommands} to the device. In recent years, the Trusted Computing Group (TCG)~\cite{TrustedComputingGroupTrustedGroup} has put a lot of effort into developing and standardizing higher level APIs~\cite{Tpm2-software:Specifications.} to interact with the TPM resulting in a toolset called \texttt{tpm2-tools}. These tools allow for higher-level interactions with the TPM using the CLI. While the low-level code complexity has been reduced significantly, a deep knowledge of the inner workings and mechanisms of a TPM is still required to actively use it in an application. Research projects like Keylime could help alleviate this complexity issue.

\subsection{Keylime}

Keylime is an open-source CNCF project originally developed by MIT's Lincoln Laboratory~\cite{Schear2016BootstrappingCloud}. It has since seen increased adoption with the support of RedHat, who are actively developing it for RHEL and OpenShift. Keylime provides another abstraction layer to TPM attestation on top of the existing \texttt{tpm2-tools} and allows developers to easily integrate boot and runtime attestation into their architectures. Keylime consists of cloud components written in Python (verifier, registrar, and tenant) and a Rust agent running on the machine that is to be attested.

\begin{figure}[h]
    \centering
    \includegraphics[width=\linewidth]{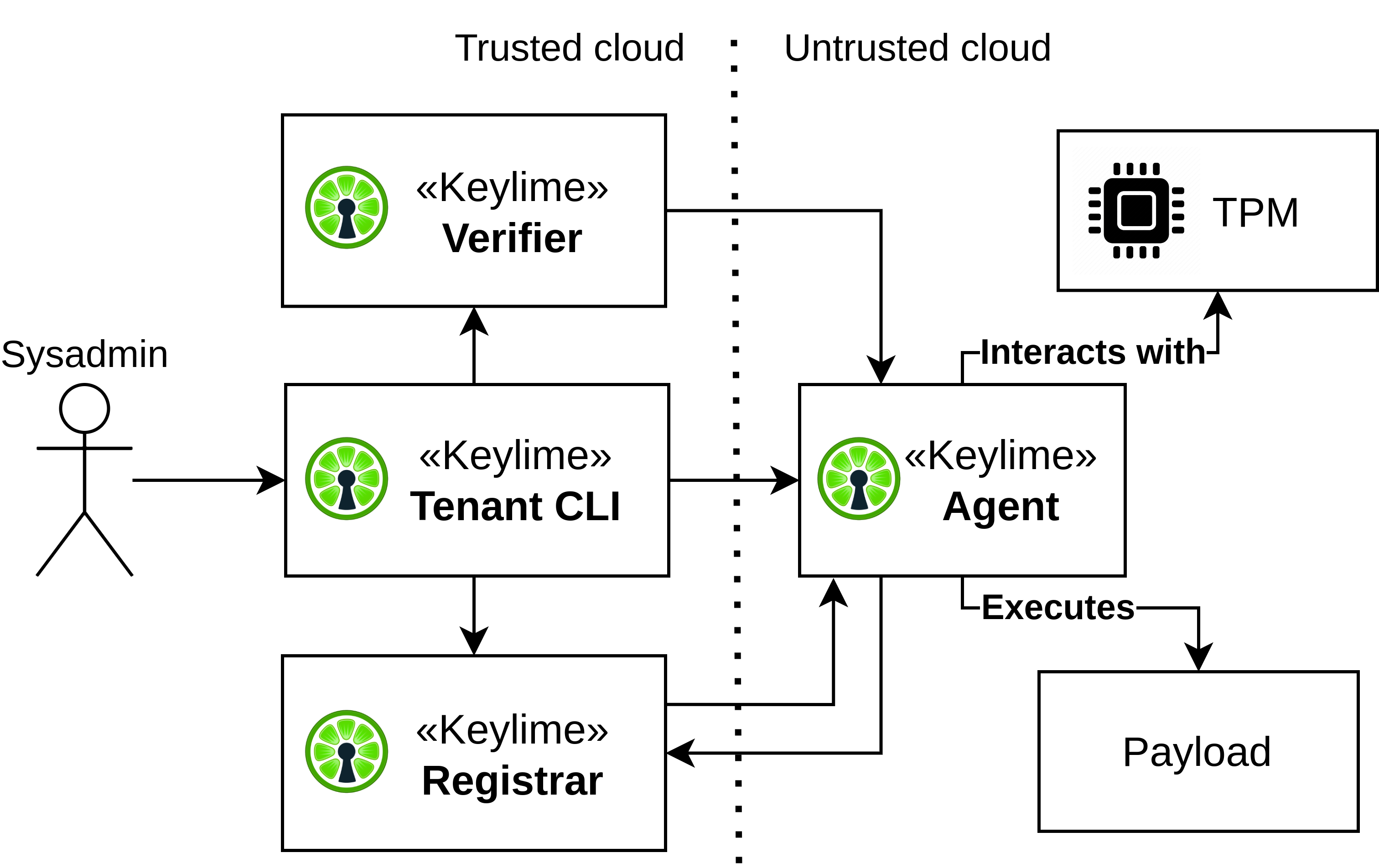}
    \caption{An overview of the Keylime architecture: a system administrator uses the Keylime tenant CLI to set up attestation of a device that runs an agent and deliver a payload.   }
    \label{fig:arch°keylime}
\end{figure}

\begin{figure*}
    \centering
    \includegraphics[width=\textwidth]{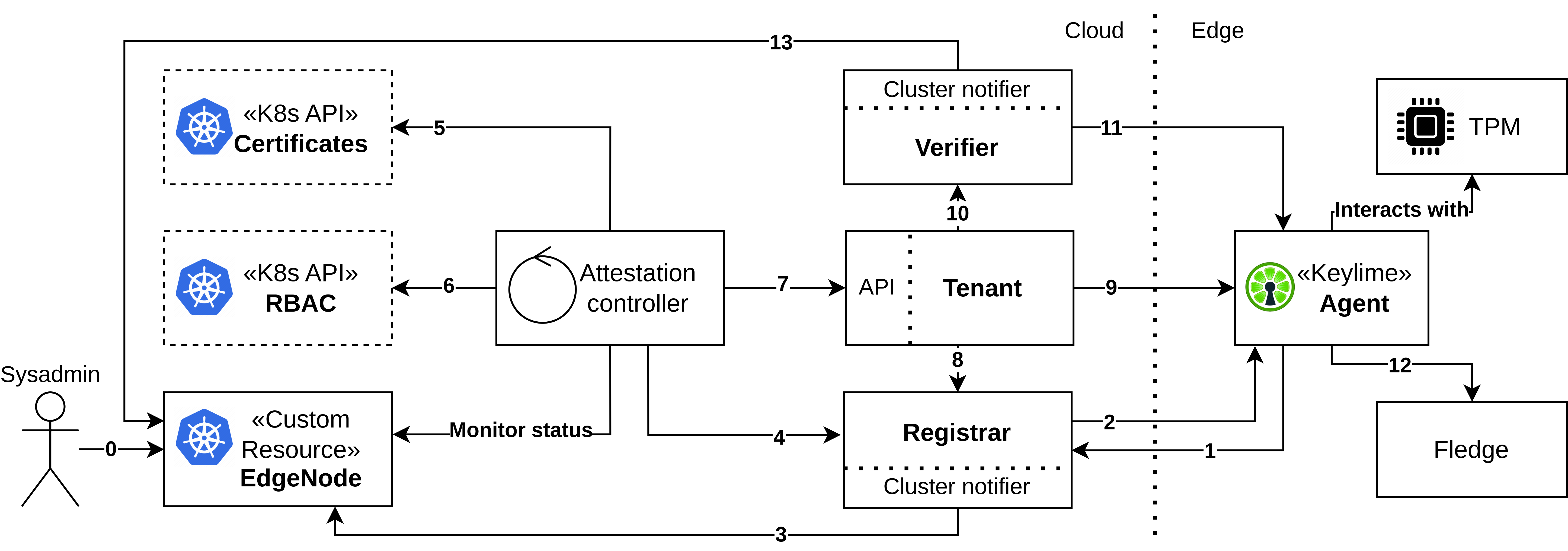}
    \caption{An overview of our architecture: the attestation controller interacts with the tenant API to enroll a trusted edge device into the cluster, it monitors the EdgeNode custom resource for attestation events and adjust RBAC permissions accordingly.}
    \label{fig:arch°overview}
\end{figure*}

The Keylime architecture is controlled by the tenant CLI application, which allows a system administrator to enroll a device and set it up for TPM-based boot and runtime attestation using specific golden values for each. These values contain known good states of the system and serve as a reference for future attestation. The tenant interacts with the other cloud components to deliver a secure payload to an agent. This agent is a device's connection point to the Keylime cloud components. It provides an API over HTTPS, which allows for an abstracted interaction with the device's TPM.

Efforts are underway to develop a cloud operator~\cite{Keylime/attestation-operator:Kubernetes/Openshift} that can manage TPM-enabled nodes in a Kubernetes compatible environment. This operator is designed to simplify the deployment of Keylime in cloud Kubernetes environments to provide node attestation. It is currently still in the early development phase. While it provides useful functionality, mainly the ability to easily deploy the complex network of Keylime cloud components using Helm, it is directed towards cloud environments and does not address the specific needs of the cloud-edge use-case outlined by this paper. Mainly the interaction with a device external to the cluster, the unique RBAC identity tied to such a device, and its dynamic RBAC permissions. For our platform we make use of an early fork of this project to deploy our customized tenant, verifier and registrar using Helm.

Berbecaru et al.~\cite{Berbecaru2022CounteractingPrototype} implemented a prototype that uses Keylime to provide attestation on a Raspberry Pi edge device and proposed some solutions to the lack of CRTM on such a device. It does however not integrate with any orchestration systems such as Kubernetes nor does it provide any automated management of the device.

\section{Architecture}\label{arch}

The next paragraphs outline the various components that compose the architecture proposed in this paper. An overview can be found in~\cref{fig:arch°overview}.

The \textbf{EdgeNode custom resource} is the center of this architecture, as it represents the edge device in the Kubernetes API. The custom resource contains all the information required to identify and attest the device, which is provided by a system administrator. The EdgeNode's status is the most important field, as it directs the controller's actions. An EdgeNode resource is initially deployed with an `unregistered' status and will move through the `registered' state towards either an `attested' or an `unattested' state throughout its lifetime. 

Kubebuilder~\cite{Kubebuilder} serves as a scaffold for the development of the \textbf{EdgeNode controller} and provides easy deployment to the cluster. The controller monitors the state of EdgeNode resources and is responsible for enrolling an edge node and steering its corresponding custom resource to an `attested state'. It interacts with the Kubernetes API to manage unique RBAC roles for each edge device and provides it with the right permissions to join the cluster. The controller interfaces with the tenant through its API to deliver these credentials to attested nodes and sets up continuous attestation. Throughout the node's lifecycle, it is responsible for modifying the unique permissions in reaction to attestation events. 

The \textbf{tenant} application is built on top of the Keylime tenant. It provides an API layer that integrates into the source of Keylime to solve shortcomings in the original design. This original design is CLI based and is meant to be called by a system administrator to enroll an attested device. The command takes a description of the node's desired state and the payload that is to be delivered as an argument. It reads these files from disk and executes the necessary operations to securely deliver the payload to an attested device. Keylime has already attempted to bypass this reliance on an interactive system administrator by offering a script that uses kubectl commands to automatically copy the necessary files to the tenant pod and execute the appropriate tenant command. This reliance on an external script however, introduces a few additional issues that can be improved upon. First, it requires multiple disk read/write operations to achieve, likely increasing latency. The controller must write the payload it generates in memory to disk, kubectl then reads it again and writes it to the tenant's disk after which the tenant can finally read it again to process it. Calling an external bash script presents a second issue, introducing extra overhead both in code complexity and code efficiency. Our API allows the tenant's functionality to be called natively by the controller using a REST interface instead of outsourcing the interactions to an external script. The API's business logic extends into the tenant's source code to carry the in-memory payload and reference boot state, generated by the controller, into the tenant to avoid expensive disk read/write operations.

Our \textbf{verifier} extends the Keylime verifier by integrating its attestation responses into the cluster. The verifier is responsible for providing continuous attestation managed by the tenant. It polls the agent for signed TPM quotes and validates them against the reference. When an attestation fails, Keylime has built in functionality to notify other components using either ZeroMQ or a webhook. To provide a better integration of the verifier into the cluster and avoid having to add extra components to the architecture, we add a third option to our verifier: a native Kubernetes revocation. By integrating the Kubernetes Python client, our verifier can patch the EdgeNode custom resource's status to reflect its unattested state. This will notify the controller, allowing it to take corrective actions.

The \textbf{registrar} plays a crucial role in storing and managing essential data for agent operation, including identity, attestation information, and metadata. Acting as the initial point of contact for edge devices, the registrar facilitates the verification of agent identity using its Endorsement Key (EK) and mTLS certificate. To automate the response to this initial contact, we extended the Keylime registrar to include Kubernetes API support. After validating the identity, it can notify the controller of the edge device's presence by patching the EdgeNode custom resource with the `registered' state, kickstarting the enrollment process.

\section{Enrollment workflow}\label{implementation}

This section covers the complete enrollment of an attested Kubernetes edge node with unique RBAC credentials. \Cref{fig:arch°overview} shows an overview of the complete flow with numbers indicating the sequence of operations.

\subsection{Registration of edge device}

First, the system administrator uses a Helm chart~\cite{Helm:Kubernetes} to create an instance of the EdgeNode custom resource~(0), loaded with the reference state and TPM identification information. The reference state consists of a JSON containing information on the allowed state of various boot components such as the CRTM, firmware, kernel, and keys. To provide identification for the TPM, an endorsement key (EK) certificate signed by the TPM's manufacturer must be included in the custom resource. An administrator can use the \texttt{tpm2-tools} to extract the EK certificate from the TPM's non-volatile memory. It is important to note that the system administrator is considered to be a fully trusted entity that determines the definition of trust for the edge device.

When an edge device is initially turned on, it will attempt to contact the registrar~(1). The registrar will subject it to an identity verification~(2) by making the device encrypt a nonce with the private EK stored in its TPM. This is validated using the matching public key contained in the EK certificate provided by the device. After successfully passing this check, the registrar adds the node's information to its records. As described in \cref{arch}, we integrated Kubernetes API support into the registrar which allows it to set the EdgeNode status to `registered'~(3), notifying the controller.

\subsection{Generation of RBAC credentials}

After device registration, the controller has to execute a series of tasks to steer the EdgeNode resource to an `attested' state and deploy the device as an edge worker node. First, the controller must verify that the entity previously registered with the registrar matches the one provided by the system administrator using the EdgeNode custom resource~(4). Simply comparing the two certificates for a match is enough since the registrar has already performed a verification of the certificate provided by the edge device. 

Next, the controller needs to provide the edge device with a unique identity and the necessary permissions to enroll as an edge worker node using an edge orchestration framework like Fledge. It generates a new key pair and uses it to request a signed certificate from the Kubernetes cluster~(5), linking the edge device to a unique Kubernetes user. This generated user does not yet have permissions to access any of the cluster resources. Next the controller creates a unique Kubernetes role for the edge device with the required RBAC permissions and links it to the user certificate using a Kubernetes rolebinding~(6). 

\subsection{Enrollment of edge node}

The payload required to enroll the edge device in the cluster is now complete. The controller packs it into a ZIP archive and together with the reference boot state from the EdgeNode custom resource, sends it to the tenant's REST endpoint~(7). The tenant encrypts the payload with a composite U-key (which is a bitwise xor of two other keys called the K- and V-keys), and starts the delivery process. First, it contacts the registrar~(8) to obtain the edge device's information, which includes its IP and port as well as the stored EK certificate. The tenant contacts the agent~(9) running on the edge device and once again verifies that the device's TPM does in fact own the EK certificate. Once the identity is established, the tenant sends the first part of the composite encryption key (U-key) as well as the encrypted payload to the edge device.

Next, the tenant forwards the second part of the payload encryption key (K-key) as well as the reference state to the verifier to set up continuous attestation~(10). The verifier requests an integrity check from the agent and validates its PCR values and boot record. After successful attestation, the verifier delivers the key to the agent~(11), allowing it to decrypt and execute the payload that starts the Fledge service and enrolls the device as an edge worker node~(12). Finally, the verifier sets the status of the EdgeNode custom resource to `attested' to notify the controller of the attestation event~(13).

The verifier will periodically check the integrity of the edge node, should a failure occur at any point it will change the EdgeNode custom resource's status to `unattested', allowing the controller to revoke RBAC permissions for the edge device's unique identity.

\section{Evaluation results}\label{eval}

We evaluated the design through three scenarios to assess its effectiveness in ensuring the integrity and security of the Kubernetes edge workers. A quantitative analysis was also performed to measure the impact of attestation on the startup time of a Kubernetes edge node. This evaluation was performed using a Dell Latitude 5540 as an edge device with 16GB RAM and an Intel Core i7. The Kubernetes cluster is provisioned using CloudNativeLab~\cite{CloudNativeLab:Testbed} with 1 worker node with 2 CPU cores and 8 GB RAM. The raw data is published on GitLab~\cite{TrustArchitecture}.

\begin{enumerate}
  \item In the first scenario, an edge device successfully passes initial attestation and receives its payload, including credentials linked to a unique RBAC role, which it uses to start Fledge and enroll as an edge node. This demonstrates the system's capability to verify the integrity of an edge device and provide them with the necessary granular credentials. The Kubernetes event log indicates a successful attestation process, with attestation being achieved 11 seconds after first contact by the edge device.
\begin{figure}
    \centering
    \includegraphics[width=\linewidth]{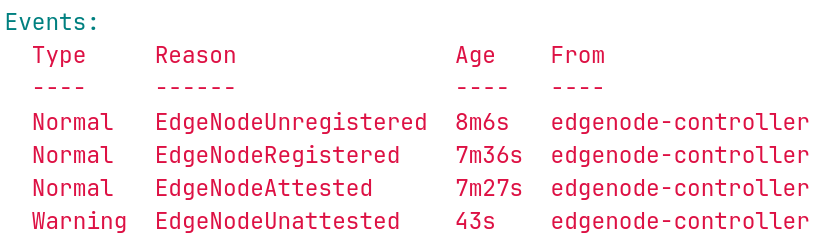}
    \caption{Kubernetes event log shows successfull detection of node attestation failure during continous monitoring.}
    \label{fig:arch°attested-unattested}
\end{figure}
%\begin{figure}
%    \centering
%    \includegraphics[width=\linewidth]{img/attested.png}
%    \caption{Kubernetes event log shows successfull attestation}
%    \label{fig:arch°attested}
%\end{figure}

    \item The second scenario involved intentionally causing a node attestation failure to evaluate the system's response. We booted a node with an unknown kernel to simulate an attacker starting a custom, possibly compromised kernel. The Kubernetes event log captures the detection of an unattested node, indicating that the system appropriately identifies integrity failures. The verifier notified the controller of this failure within 7 seconds. No payload was delivered to the edge device and its RBAC permissions were taken away, denying it the ability to enroll as a worker node. 

%\begin{figure}
%    \centering
%    \includegraphics[width=\linewidth]{img/unattested.png}
%    \caption{Kubernetes event log shows successfull detection of an unattested node}
%    \label{fig:arch°unattested}
%\end{figure}

    \item In the third scenario, the system's continuous monitoring capability was assessed by inducing a node attestation failure after the system had already been successfully attested. This simulates an attacker rebooting an attested system that already serves as a worker node and loading a compromised kernel. \cref{fig:arch°attested-unattested} illustrates the Kubernetes event log detecting a node attestation failure during continuous monitoring. In response to this failure, the controller swiftly retracted the RBAC permissions linked to this worker's unique identity, preventing it from accessing cluster resources. This scenario validates the system's effectiveness in maintaining security posture over time by promptly identifying attestation failure and revoking cluster access for compromised nodes.

\end{enumerate}

\begin{figure}[h]
    \centering
    \includegraphics[width=\linewidth]{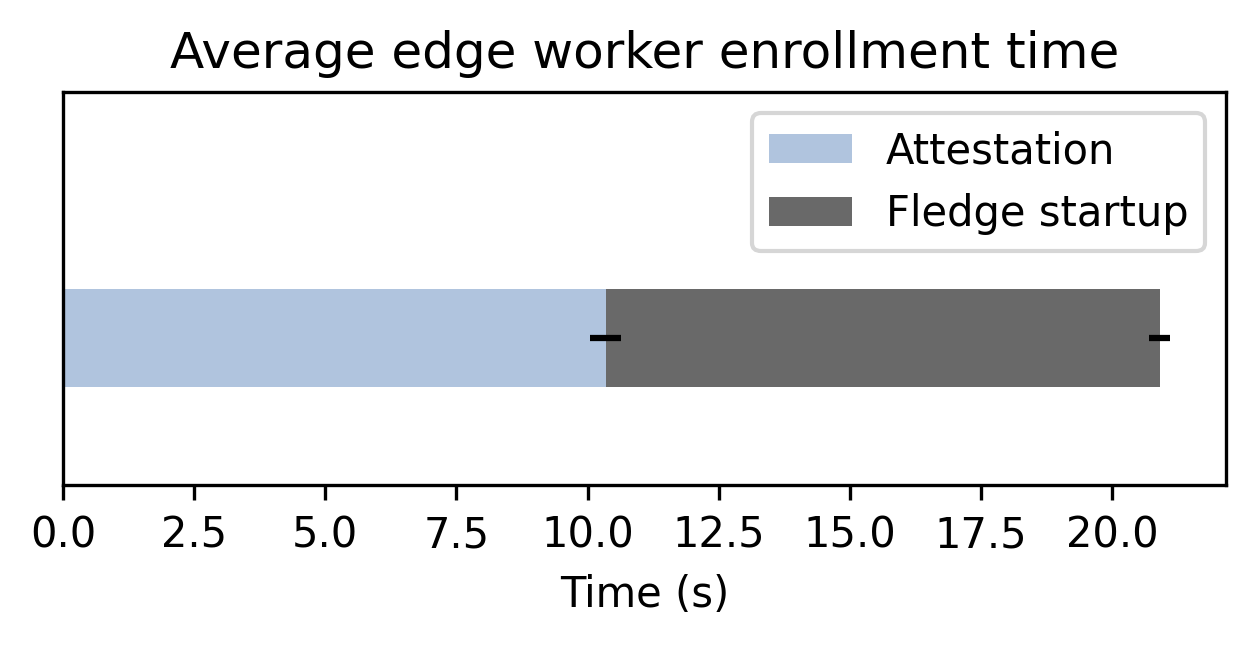}
    \caption{Attestation takes on average 10.28 seconds for a total node registration time of 20.91 seconds.}
    \label{fig:graph}
\end{figure}

Finally, we present a quantitative evaluation (\cref{fig:graph}). We executed the first scenario 50 times and measured the time between the agent service and the Fledge service starting as well as the time Fledge needs to register as an edge worker. The results indicate an increase in startup time due to attestation of 10.28 seconds (std 0.30s). This is in addition to the startup time of Fledge, which measured on average, 10.52 seconds (std 0.20s), for a total average startup time of an attested edge node of 20.91 seconds (std 0.36s). %An overview of these result can be seen in~\cref{fig:graph}.

The evaluation results of the proposed architecture effectively address the research questions. Scenario one demonstrates the architecture's capability in securely enrolling edge nodes into a Kubernetes cluster (\textbf{RQ1}) by verifying device integrity and delivering necessary credentials. Furthermore, scenario two and three show that the system provides granular cluster permissions in response to attestation events (\textbf{RQ2}), dynamically adjusting RBAC permissions based on node status. Thus, the evaluation validates the system's ability to achieve secure enrollment and provide tailored permissions, aligning with the research objectives.
\section{Future work}\label{future-work}
A major challenge remains the \textbf{binary nature of attestation}. In traditional schemes, a node either passes or fails attestation. When using boot attestation, this means that a node running a newer kernel that was not included in the reference will automatically be considered insecure. This binary decision is at odds with the flexible and granular nature of Kubernetes access control and scheduling. Instead of deciding on a pass/fail basis if a certain combination of CRTM, firmware, and kernel should be trusted, a more nuanced description of trust must be employed. An older kernel version is not necessarily insecure, but to comply with security SLAs one could limit such a node's access to sensitive resources and prevent the scheduling of certain workloads to such nodes while still allowing it to execute non-sensitive ones.

A final problem that is outside the scope of this paper but requires future attention is \textbf{network access control}. This paper focuses on setting up a unique RBAC identity for an attested edge node that allows flexible adjustments to Kubernetes resource access. It does however, not cover that node's access to the different networks. The node currently has full access to the local network through its VPN connection, and thus any resources exposed on any of the cloud nodes. This network access should also be managed by the controller, initially limiting the connection to only the registrar and expanding network access as the attestation state evolves.

\section{Conclusion}

In conclusion, this research paper proposes a novel architecture and open-source implementation for secure and automated enrollment of Kubernetes edge nodes after remote attestation, ensuring the integrity of the entire cluster even in physically exposed environments. By leveraging hardware-based boot attestation rooted in Trusted Platform Modules (TPMs), the architecture establishes a strong base of trust for edge devices. The integration of role-based access control (RBAC) mechanisms enables granular adjustments to access permissions in response to attestation events, further protecting sensitive cluster resources from untrusted edge nodes.

Through qualitative scenarios and a quantitative evaluation, the effectiveness of the proposed architecture has been demonstrated. The system successfully enrolls edge nodes into the Kubernetes cluster, verifies their integrity, and dynamically adjusts RBAC permissions based on attestation events. Attestation adds on average 10.28 seconds to the enrollment time, for a total average of 20.91 seconds. This addresses the research questions regarding automated enrollment and integration with authentication and authorization frameworks, showcasing the architecture's capability in enhancing security within cloud-edge computing ecosystems.

We identify future work in addressing the binary nature of attestation decisions and implementing network access control to further bolster the security of edge deployments. Overall, this research paves the way for resilient and secure edge computing environments, crucial for the proliferation of distributed and decentralized computing infrastructures.

\bibliographystyle{IEEEtran}
\bibliography{references}

\begin{thebibliography}{10}
\providecommand{\url}[1]{#1}
\csname url@samestyle\endcsname
\providecommand{\newblock}{\relax}
\providecommand{\bibinfo}[2]{#2}
\providecommand{\BIBentrySTDinterwordspacing}{\spaceskip=0pt\relax}
\providecommand{\BIBentryALTinterwordstretchfactor}{4}
\providecommand{\BIBentryALTinterwordspacing}{\spaceskip=\fontdimen2\font plus
\BIBentryALTinterwordstretchfactor\fontdimen3\font minus
  \fontdimen4\font\relax}
\providecommand{\BIBforeignlanguage}[2]{{%
\expandafter\ifx\csname l@#1\endcsname\relax
\typeout{** WARNING: IEEEtran.bst: No hyphenation pattern has been}%
\typeout{** loaded for the language `#1'. Using the pattern for}%
\typeout{** the default language instead.}%
\else
\language=\csname l@#1\endcsname
\fi
#2}}
\providecommand{\BIBdecl}{\relax}
\BIBdecl

\bibitem{Kubernetes:Applications}
\BIBentryALTinterwordspacing
``{Kubernetes: open-source system for automating deployment, scaling, and
  management of containerized applications}.'' [Online]. Available:
  \url{https://kubernetes.io/}
\BIBentrySTDinterwordspacing

\bibitem{Goethals2022ExtendingKubelets}
T.~Goethals, F.~De~Turck, and B.~Volckaert, ``{Extending Kubernetes Clusters to
  Low-Resource Edge Devices Using Virtual Kubelets},'' \emph{IEEE Transactions
  on Cloud Computing}, vol.~10, no.~4, pp. 2623--2636, 10 2022.

\bibitem{KubeEdge:Framework}
\BIBentryALTinterwordspacing
``{KubeEdge: a Kubernetes Native Edge Computing Framework}.'' [Online].
  Available: \url{https://kubeedge.io/}
\BIBentrySTDinterwordspacing

\bibitem{Goel2022AuthenticatingCluster}
A.~Goel and B.~Thangaraju, ``{Authenticating Distributed Systems Using SPIRE
  over Kubernetes Cluster},'' in \emph{2022 IEEE International Conference on
  Electronics, Computing and Communication Technologies (CONECCT)}.\hskip 1em
  plus 0.5em minus 0.4em\relax IEEE, 7 2022, pp. 1--6.

\bibitem{Fernandez2019SecureCloud}
G.~P. Fernandez and A.~Brito, ``{Secure container orchestration in the
  cloud},'' in \emph{Proceedings of the 34th ACM/SIGAPP Symposium on Applied
  Computing}.\hskip 1em plus 0.5em minus 0.4em\relax New York, NY, USA: ACM, 4
  2019, pp. 138--145.

\bibitem{Keylime:IoT}
\BIBentryALTinterwordspacing
``{Keylime: Bootstrap {\&} Maintain Trust on the Edge / Cloud and IoT}.''
  [Online]. Available: \url{https://keylime.dev/}
\BIBentrySTDinterwordspacing

\bibitem{IslamShamim2020XIPractices}
M.~S. Islam~Shamim, F.~Ahamed~Bhuiyan, and A.~Rahman, ``{XI Commandments of
  Kubernetes Security: A Systematization of Knowledge Related to Kubernetes
  Security Practices},'' in \emph{2020 IEEE Secure Development (SecDev)}.\hskip
  1em plus 0.5em minus 0.4em\relax IEEE, 9 2020, pp. 58--64.

\bibitem{TrustKubernetes}
\BIBentryALTinterwordspacing
``{Trust Edge: Trusted edge nodes for Kubernetes}.'' [Online]. Available:
  \url{https://github.com/idlab-discover/trust-edge}
\BIBentrySTDinterwordspacing

\bibitem{Kinney20066Registers}
\BIBentryALTinterwordspacing
S.~Kinney, ``{6 - Platform Configuration Registers},'' in \emph{Trusted
  Platform Module Basics}, ser. Embedded Technology, S.~Kinney, Ed.\hskip 1em
  plus 0.5em minus 0.4em\relax Burlington: Newnes, 2006, pp. 53--64. [Online].
  Available:
  \url{https://www.sciencedirect.com/science/article/pii/B9780750679602500075}
\BIBentrySTDinterwordspacing

\bibitem{CooperDavid2011SpecialGuidelines}
{Cooper David}, {Polk William}, {Regenscheid Andrew}, and {Souppaya Muragiah},
  ``{Special Publication 800- 147: BIOS Protection Guidelines},'' NIST,
  Gaithersburg, Tech. Rep., 4 2011.

\bibitem{Arthur2015A2.0}
W.~Arthur, D.~Challener, and K.~Goldman, ``{A Practical Guide to TPM
  2.0}.''\hskip 1em plus 0.5em minus 0.4em\relax Apress, 2015, p. 152.

\bibitem{Raj2016FTPM:Chip}
\BIBentryALTinterwordspacing
H.~Raj, S.~Saroiu, A.~Wolman, R.~Aigner, J.~Cox, P.~England, C.~Fenner,
  K.~Kinshumann, J.~Loeser, D.~Mattoon, M.~Nystrom, D.~Robinson, R.~Spiger,
  S.~Thom, and D.~Wooten, ``{fTPM: A Software-Only Implementation of a TPM
  Chip},'' in \emph{25th USENIX Security Symposium (USENIX Security 16)}.\hskip
  1em plus 0.5em minus 0.4em\relax Austin, TX: USENIX Association, 8 2016, pp.
  841--856. [Online]. Available:
  \url{https://www.usenix.org/conference/usenixsecurity16/technical-sessions/presentation/raj}
\BIBentrySTDinterwordspacing

\bibitem{Sun2018ETPM:Technology}
\BIBentryALTinterwordspacing
H.~Sun, R.~He, Y.~Zhang, R.~Wang, W.~H. Ip, and K.~L. Yung, ``{eTPM: A Trusted
  Cloud Platform Enclave TPM Scheme Based on Intel SGX Technology},''
  \emph{Sensors}, vol.~18, no.~11, 2018. [Online]. Available:
  \url{https://www.mdpi.com/1424-8220/18/11/3807}
\BIBentrySTDinterwordspacing

\bibitem{2019TrustedArchitecture}
``{Trusted Platform Module Library Part 1: Architecture},'' Trusted Computing
  Group, Tech. Rep., 11 2019.

\bibitem{2019TrustedCommands}
``{Trusted Platform Module Library Part 3: Commands},'' Trusted Computing
  Group, Tech. Rep., 11 2019.

\bibitem{TrustedComputingGroupTrustedGroup}
{Trusted Computing Group}, ``{Trusted Computing Group},''
  https://trustedcomputinggroup.org/.

\bibitem{Tpm2-software:Specifications.}
\BIBentryALTinterwordspacing
``{tpm2-software: Developer community for those implementing APIs and
  infrastructure from the TCG TSS2 specifications.}'' [Online]. Available:
  \url{https://github.com/tpm2-software}
\BIBentrySTDinterwordspacing

\bibitem{Schear2016BootstrappingCloud}
N.~Schear, P.~T. Cable, T.~M. Moyer, B.~Richard, and R.~Rudd, ``{Bootstrapping
  and maintaining trust in the cloud},'' in \emph{Proceedings of the 32nd
  Annual Conference on Computer Security Applications}.\hskip 1em plus 0.5em
  minus 0.4em\relax New York, NY, USA: ACM, 12 2016, pp. 65--77.

\bibitem{Keylime/attestation-operator:Kubernetes/Openshift}
\BIBentryALTinterwordspacing
``{keylime/attestation-operator: Keylime easily deployable on
  Kubernetes/Openshift}.'' [Online]. Available:
  \url{https://github.com/keylime/attestation-operator}
\BIBentrySTDinterwordspacing

\bibitem{Berbecaru2022CounteractingPrototype}
D.~G. Berbecaru and S.~Sisinni, ``{Counteracting software integrity attacks on
  IoT devices with remote attestation: a prototype},'' in \emph{2022 26th
  International Conference on System Theory, Control and Computing
  (ICSTCC)}.\hskip 1em plus 0.5em minus 0.4em\relax IEEE, 10 2022, pp.
  380--385.

\bibitem{Kubebuilder}
\BIBentryALTinterwordspacing
``{kubebuilder}.'' [Online]. Available: \url{https://book.kubebuilder.io/}
\BIBentrySTDinterwordspacing

\bibitem{Helm:Kubernetes}
\BIBentryALTinterwordspacing
``{Helm: The package manager for Kubernetes}.'' [Online]. Available:
  \url{https://helm.sh/}
\BIBentrySTDinterwordspacing

\bibitem{CloudNativeLab:Testbed}
\BIBentryALTinterwordspacing
``{CloudNativeLab: Kubernetes testbed}.'' [Online]. Available:
  \url{https://practicum.cloudnativelab.ilabt.imec.be/}
\BIBentrySTDinterwordspacing

\bibitem{TrustArchitecture}
\BIBentryALTinterwordspacing
``{Trust Benchmark: Benchmark results for trust edge architecture}.'' [Online].
  Available: \url{https://gitlab.ilabt.imec.be/edge-keylime/trust-benchmark}
\BIBentrySTDinterwordspacing

\end{thebibliography}
\end{document}